\documentclass[preprint,nofootinbib,aps,amsfonts,superscriptaddress]{revtex4}
\usepackage{amsfonts,amssymb,amscd,amsmath}
\usepackage{graphicx}
\usepackage{helvet}  
\usepackage[T1]{fontenc}  
\usepackage[latin1]{inputenc}  
\usepackage{graphicx}  
\usepackage{amssymb}  

\newcommand{\slashit}[1]{#1 \kern-.45em\slash}

\newcommand{\slashP}{P \kern-.65em\slash }
\begin{document}
\title{\Large Charmed Meson Production in Deep Inelastic Scattering}  
\author{B.~Fl\"oter}
\affiliation{Institute for Theoretical Physics, University of
  Heidelberg, Philosophenweg 19, 69120 Heidelberg, Germany}
\author{ B. Z. Kopeliovich}
\affiliation{Departamento de F\'\i sica y Centro de Estudios Subat\'omicos, Universidad
T\'ecnica Federico Santa Mar\'\i a, Casilla 110-V, Valpara\'\i so,
Chile}
\affiliation{Joint Institute for Nuclear Research, Dubna, Russia}
\author{H.-J.~Pirner}
\affiliation{Institute for Theoretical Physics, University of
Heidelberg, Philosophenweg 19, 69120 Heidelberg, Germany}
\affiliation{Max-Planck-Institut f\"ur Kernphysik, Saupfercheckweg 1,
69117 Heidelberg, Germany}
\author{J.~Raufeisen}
\affiliation{Institute for Theoretical Physics, University of
Heidelberg, Philosophenweg 19, 69120 Heidelberg, Germany}

\providecommand{\LyX}{L\kern-.1667em\lower.25em\hbox{Y}\kern-.125emX\@}  
\newcommand{\noun}[1]{\textsc{#1}}  
  
\newcommand{\Tr}{\mathrm{Tr}}  
\newcommand{\tr}{\mathrm{tr}}  
\newcommand{\dd}{\displaystyle}  
\newcommand{\nn}{\nonumber}  
\newcommand{\be}{\begin{eqnarray}}  
\newcommand{\ee}{\end{eqnarray}}  
\newcommand{\bea}{\begin{eqnarray}}  
\newcommand{\eea}{\end{eqnarray}}  
  
\def\del{\partial\kern -0.53em/}  
\def\Del{D\kern -0.63em/}  
\def\SlashP{p\kern -0.43em/}  
\def\Tr{\mathrm{Tr}}  
\def\tr{\mathrm{tr}}  
\def\Det{\mathrm{Det}}  
\def\det{\mathrm{det}}  
\def\itop{\mathrm{T}}

\date{\today}  
  
\vspace*{2cm}
  
\begin{abstract}  
  
  Charmed meson production in semi-inclusive deep inelastic scattering  
  is investigated in the color dipole formalism. The transverse  
  momentum distributions are calculated. We find good agreement with  
  the H1 data using a hard fragmentation function.  
  
\end{abstract}  
  
\maketitle  
  
\section{Introduction}  
  
One of the most important heavy-flavor production processes is deep
inelastic scattering (DIS). A quantitative understanding of heavy-flavor
physics in $ep$ and $pp$-collisions is a prerequisite for the discovery of
new effects in $pA$ and $AA$-collisions.  The color dipole model represents
a good phenomenological approach to heavy-flavor production.  In this
model, the DIS cross section is factorized into a light-cone wavefunction,
which describes the splitting of the virtual photon into a colorless
quark-antiquark dipole with transverse separation $r$ and a dipole cross
section which depends on $r$ and describes subsequent dipole scattering off
the proton. The dipoles are eigenstates of the interaction and thus,
multiple scattering effects on nuclei are easily described \cite{zkl}.
Historically, this was the initial intention to develop the dipole
formalism. The dipole approach can be formulated in the target rest frame.
For low-$x$ the typical propagation length of a dipole fluctuation exceeds
the interaction time by orders of magnitude. Because of time dilatation we
can think of a frozen transverse separation during the interaction with the
target. It is expected that the growth of the gluon density is slowed down
at very low $x_{B}$ by gluon-gluon recombination. Within the dipole
approach, this saturation effect is described most naturally by introducing
an $x$-dependent saturation scale. A fast-moving dipole produced from the
incident virtual photon decouples from soft QCD interactions. This color
transparency manifests itself in the dependence of the dipole cross section
proportional to $r^2$ for small dipoles. The color dipole approach is
inherently nonperturbative and includes higher twist effects which are
important at low x-Bjorken and for low transverse momenta of the produced
D-mesons.
  
The paper is organized as follows. Section 2 deals with $D$-meson  
production in deep inelastic scattering. We develop the kinematic  
framework for the description of semi-inclusive deep inelastic  
scattering. Section 3 introduces the dipole approach to DIS and  
presents the virtual photon-proton cross section preparing the  
numerical evaluation of the deep inelastic cross sections. In section 4 we   
finalize the dipole calculation and compare with data from HERA's H1  
collaboration.  Finally, in section 5 we summarize the results.

\section{Semi-inclusive $D$-meson production in deep inelastic scattering}  
  
In inclusive Deep Inelastic Scattering (DIS) we consider the process $l\,N\rightarrow  
l'\,X$, where a lepton $l$ with momentum $k$ scatters off the nucleon  
$N$ with momentum $P$ resulting in a momentum $k'$ for the scattered  
lepton $l'$. The target breaks up into an unobserved final state $X$.  
The cross section is most conveniently expressed in terms of the  
Lorentz invariants  
\begin{equation}  
Q^{2}=-q^{2}=-(k-k')^{2}>0 \qquad \nu=\frac{P\cdot q}{M}  
\end{equation}  
or the ratios  
\begin{equation}  
x_{B}=\frac{Q^{2}}{2M\nu}\qquad y=\frac{P\cdot q}{P\cdot k} . 
\end{equation}  
In the target rest frame, i.e. in the rest frame of the nucleon, $\nu$  
gives the energy transfer from the lepton to the nucleon and $y$ the  
ratio of energy transfer to incident lepton energy. $Q^{2}$ is the  
virtual mass squared (virtuality) of the exchanged photon. We also use  
the center-of-mass energy squared in the $l\,N$ system $s$ and in the  
$\gamma^{*}p$ system 
\begin{equation}  
s=(k+P)^{2}, \qquad W^{2}=(q+P)^{2}.  
\end{equation}  
  
\begin{figure}[htbp]  
\begin{center}  
\resizebox{12cm}{!}{\includegraphics{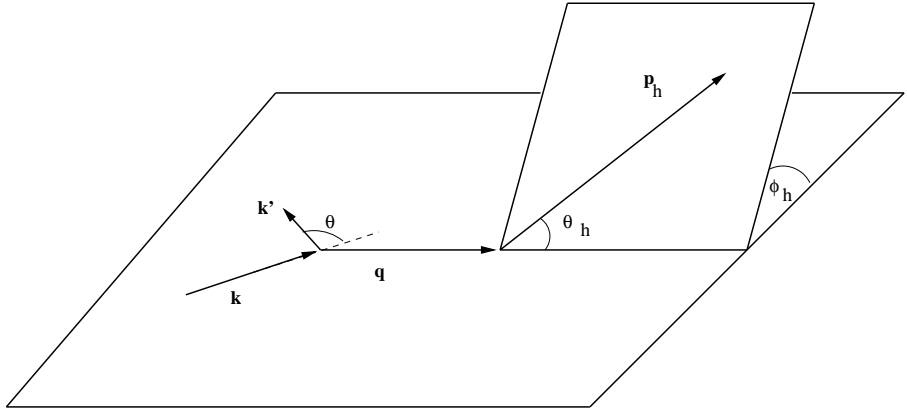}}  
\caption{Kinematics of one-particle semi-inclusive scattering. The  
  reaction plane in $(e,e'h)$-scattering is defined by the initial and  
  final momenta of the lepton $(\vec k,\vec k')$. The produced hadron  
   $\vec p_h$ and photon momenta $\vec q$ define a second plane.}  
\label{fig:frame}  
\end{center}  
\end{figure}

In Semi-inclusive Deep Inelastic Scattering (SIDIS) \cite{Levelt:1993ac,le}   
of the type $lH\rightarrow l'hX$  
the hadron $h$ is   detected in coincidence with the scattered  
electron $e'$ in the final state.   
The corresponding situation is pictured in Fig.~\ref{fig:frame}. The hadronic target 
$H$ has mass $M$ and  
four-momentum $P$, the produced hadron $h$ has mass $m$ and  
four-momentum $p_h$.  
  
The SIDIS cross section can be expressed in the same Lorentz invariants  
$x_{B}$ and $y$ as for inclusive DIS plus the final state hadron  
variables $p_{h}$. Since we aim for $p_{T}$ factorization we split the  
final state hadron variables into the Lorentz invariant $z$  
momentum fraction of the hadron relative to the struck quark  
\begin{equation}  
z=\frac{P\cdot p_{h}}{P\cdot q}  
\label{eqn:z}  
\end{equation}  
and the transverse hadron momenta $p_{h_\perp}$ relative to the  
$\gamma^{*}$ momentum direction. Both are invariant under boosts along  
the virtual photon axis. The differentials are related by  
\begin{equation}  
\frac{d^{3}p_{h}}{2E_{h}}=\frac{1}{4}
\frac{\nu}{p_{h\vert\vert}}dzdp_{h\bot}^{2}d\phi_{h} . 
\label{eqn:differentials}  
\end{equation}  
In analogy to inclusive DIS the SIDIS cross section can be  
obtained from the contraction of the hadronic and the leptonic tensor  
\begin{eqnarray}  
\frac{2E_{h}\,d\sigma}{d^{3}p_{h}\,dx_{B}\,dy}  
&=&\frac{\pi\alpha^{2}2My}{Q^{4}}L_{\mu\nu}W^{\mu\nu}\nonumber\\  
\frac{2E_{h}\,d\sigma}{d^{3}p_{h}\,d\Omega\, dE'}
&=&\frac{\alpha^{2}}{Q^{4}}L_{\mu\nu}W^{\mu\nu}.  
\label{eqn:sidistensors}  
\end{eqnarray}  
  
The conversion between these two forms is given by the Jacobian
\begin{displaymath}
dx_Bdy=\frac{E'}{M(E-E')}dE'd\cos \theta
\end{displaymath}
for azimuthal symmetry. $L_{\mu\nu}$ is the spin averaged leptonic
tensor and $W_{\mu\nu}$ is the hadronic tensor
\begin{eqnarray}
W_{\mu\nu}&=&\frac{1}{2M(2\pi)^{4}}
\int\frac{d^{3}P_{X}}{(2\pi)^{3}2E_{X}}<P\vert J_{\nu}(0)
\vert P_{X},p_{h}><p_{h},P_{X}\vert J_{\mu}(0)\vert P>\nonumber\\
&\times&(2\pi)^{4}\delta^{4}(P+q-p_{h}-P_{X}).
\label{eqn:sihadrtens}
\end{eqnarray}  
The semi-inclusive hadronic tensor is related to the inclusive one via
\begin{equation}
<n_{h}(P,q)>W_{\mu\nu}(P,q)=\int\frac{d^{3}p_{h}}{2E_{h}}W_{\mu\nu}(P,q,p_{h})
\label{eqn:htrel}
\end{equation}
where $<n_{h}(P,q)>$ is the average number of particles produced of 
type $h$ in a reaction defined by $P$ and $q$. The above notation
underlines the dependence on the momenta of the deep inelastic
reaction. Integrating over the hadron momentum $p_{h}$ one recovers
the inclusive hadronic tensor times the multiplicity as shown on the right 
hand
side of Eq. 8. Summation over all possible
hadrons $h$ would lead to the inclusive hadronic tensor.  The
tensorial structure of $W_{\mu,\nu}$ is determined by Lorentz and
gauge invariance.\\
Assuming time reversal and parity invariance we represent the     
electromagnetic hadronic tensor with four structure functions
$W_{i}(x,Q^{2},z,\vec p_{h}^{2})$\quad\cite{Levelt:1993ac}
\begin{eqnarray}
W_{\mu\nu}(P,p_{h},q)
&=&(\frac{q_{\mu}q_{\nu}}{q^{2}}-g_{\mu\nu})W_{1}+\frac{T_{\mu}T_{\nu}}{M^{2}}W_{2}\nonumber\\
&+&\frac{p_{h\mu}^{\bot}T_{\nu}+T_{\mu}p_{h\nu}^{\bot}}
{Mm_{h}}W_{3}+\frac{p_{h\mu}^{\bot}p_{h\nu}^{\bot}}{m_{h}^{2}}W_{4}
\label{eqn:sihtpara}
\end{eqnarray}
where
\begin{equation}
T^{\mu}=P^{\mu}-\frac{p\cdot q}{q^{2}}q^{\mu}.
\end{equation}
The first two structure functions have the same origin as in the
inclusive structure functions which are related to longitudinal and
transverse photon scattering. The dependence on the hadronic angle
$\phi_h$ allows two more structure functions. Next we introduce four
dimensionless semi-inclusive structure functions
$H_{i}(x,Q^{2},z,p_{h\bot}^{2})$. A factor of $2z$  arises in formal
analogy to the known inclusive structure functions $F_{1,2}$
\begin{eqnarray}
2zH_{1}&=&M(W_{1}+\frac{p_{h\bot}^{2}}{2m_{h}^{2}}W_{4})\nonumber\\
2zH_{2}&=&\nu (W_{2}+\frac{p_{h\bot}^{2}Q^{2}}{2m_{h}^{2}\vec q^{2}}W_{4}).
\label{eqn:h's}
\end{eqnarray}
The integration over the azimuthal angle $\phi_{h}$ removes the 
$H_{3}$ and $H_{4}$ structure functions from the semi-inclusive cross 
section
\begin{eqnarray}
H_{i}(x_{B},Q^{2},z)&=&
\frac{1}{2}\int 
dp_{h\bot}^{2}d\phi_{h}H_{i}(x_{B},Q^{2},z,p_{h\bot}^{2})\nonumber\\
&=&\pi\int dp_{h\bot}^{2}H_{i}(x_{B},Q^{2},z,p_{h\bot}^{2})
\end{eqnarray}
which leads to
\begin{equation}
\frac{d\sigma}{dx_{B}\,dz\,dy}=
\left.\frac{8\pi\alpha^{2}ME}{Q^{4}}
\right[x_{B}y^{2}H_{1}(x_{B},Q^{2},z,p_{h\bot}^{2})+(1-y)H_{2}(x_{B},Q^{2},z,p_{h\bot}^{2})\left.
\vphantom{\frac{1}{1}}\right].
\label{eqn:sidisanalogie}
\end{equation}
In the high-energy approximation $E=s/(2M)$ the cross section has a
form similar to the inclusive cross section. Factorization of deep
inelastic scattering and hadronization allows to represent the
inclusive structure function $H_2$ as a product of the known quark
distribution functions and the quark fragmentation function
$D_{h/q_{f}}(z)$ \cite{be}
\begin{equation}
H_{2}(x_{B},Q^{2},z)=H_{2}(x_{B},z)=
\sum_{f}e_{f}^{2}[q_{f/H}(x_{B})D_{h/q_{f}}(z)+\bar q_{f/H}(x_{B})D_{h/\bar 
q_{f}}(z)].
\end{equation} 
This factorization has been proven to leading order in $Q$ \cite{le}.
Here, in this paper, we want to calculate the structure functions
$H_i$ in the dipole formalism for $\gamma^{*}p$.  Using equation
(\ref{eqn:htrel}), we connect the $H_{i}$ to the standard structure
functions $F_{i}$ for inclusive lepton-hadron scattering,
\begin{eqnarray}
<n_{h}(x_{B},Q^{2})>F_{1}(x_{B},Q^{2})&=&\int 
dz\,H_{1}(x_{B},z,Q^{2})\nonumber\\
<n_{h}(x_{B},Q^{2})>F_{2}(x_{B},Q^{2})&=&\int dz\,H_{2}(x_{B},z,Q^{2}).
\end{eqnarray}
Following the inclusive case, we need two structure functions
specified for longitudinal and transverse photon polarization.
Polarization interference in semi-inclusive DIS occurs only for $W_3$
and $W_4$ which vanish after integration over the azimuthal angle.
\begin{eqnarray}
W_{L}&=&
\varepsilon(\lambda=0)\cdot W\cdot \varepsilon(\lambda=0)=
2z\left(\frac{\vec q^{2}}{\nu Q^{2}}H_{2}-\frac{1}{M}\right)\\
W_{T}&=&
\frac{1}{2}\left[\varepsilon(\lambda=-1)\cdot 
W\cdot\varepsilon(\lambda=-1)+
\varepsilon(\lambda=+1)\cdot 
W\cdot\varepsilon(\lambda=+1)\right]=\frac{2z}{M}H_{1}\nonumber
\end{eqnarray}
where $\varepsilon$ is the virtual photon polarization vector. For the
virtual photon momentum $q=(q^{0},0,0,q^{z})$ we define the polarization 
vectors:
\begin{equation}
\varepsilon(\lambda=0)=
\frac{1}{Q}\left(q^{z},0,0,q^{0}\right) \qquad \varepsilon(\lambda=\pm 1)=
\frac{1}{\sqrt{2}}(0,1,\pm i,0).
\label{eqn:polvector}
\end{equation}
The combined longitudinal and transverse electron-proton cross section is
\begin{equation}
\frac{d\sigma}{dx_{B}\,dz\,dy\,dp_{h\bot}^{2}\,d\phi_{h}}=
\frac{\pi\alpha^{2}M}{2Q^{2}yz}\left\lbrace 
\left(\frac{y^{2}}{2}+1-y\right)
W_{T}+(1-y)W_{L}\right\rbrace.\label{17}\\
\end{equation}
We recall the relation between the total $\gamma^{*}p$ cross section
and the inclusive hadronic tensor $W^{\mu\nu}(P,q)$
\begin{equation}
\sigma^{tot}_{\gamma^{*}p\rightarrow X}(\lambda)=
\frac{16\pi^{2}\alpha M}{2ys}\varepsilon^{*}_{\mu}(\lambda)
W^{\mu\nu}(P,q)\varepsilon_{\nu}(\lambda).
\end{equation}
Since the amount of $D$-meson production is  measured by the branching 
ratio
\begin{equation}
<n_{D}>=\frac{1}
{\sigma^{tot}_{\gamma^{*}p\rightarrow X}(\lambda)}
\int dy\, \frac{\partial\sigma^{inc}_{\gamma^{*}p\rightarrow 
D\,X}(\lambda)}{\partial y}
\end{equation}  
 we obtain
\begin{equation}
\frac{2E_{D}\,d\sigma^{\gamma^{*}p\rightarrow DX}}{d^{3}p_{D}}=
\frac{8\pi^{2}\alpha M}{ys}\varepsilon^{*}_{\mu}(\lambda)
W^{\mu\nu}(P,q,p_{D})\varepsilon_{\nu}(\lambda),
\end{equation}
or with (\ref{eqn:differentials})
\begin{equation}
\frac{ysz}{\pi^{2}M\alpha_{em}}\frac{d\sigma^{L,T}}{d^2{p_{D}}\,dz}=
\varepsilon^{L,T}\cdot W\cdot \varepsilon^{L,T}.
\end{equation}
We insert this expression in (\ref{17}) to find
\begin{eqnarray}
\frac{d\sigma(ep\rightarrow DX)}{dx\,dy\,dz\,d^{2}p_{D}^{\bot}}&=
&\left.\frac{\alpha_{em}}{\pi xy}\right[\left(\frac{y^{2}}{2}-y+1\right)
\frac{d\sigma^{T}(\gamma^{*}p\rightarrow 
DX)}{dz\,d^{2}p_{D}^{\bot}}\nonumber\\
&+&(1-y)\left.\frac{d\sigma^{L}(\gamma^{*}p\rightarrow 
DX)}{dz\,d^{2}p_{D}^{\bot}}\right].
\label{eqn:epges}
\end{eqnarray}

This equation relates the SIDIS cross section to the semi-inclusive cross
section of a virtual photon on the proton. The latter can easily be
calculated in the color dipole model, since the photon is the source of the
color dipole, and the target proton mainly delivers the low x-gluon which
is necessary for the virtual heavy quark-antiquark state in the photon to
materialize.
  
\section{ $D$-Meson production in the color-dipole model}  
\label{sec:dipmod}  

\begin{figure}[htbp]  
\begin{center}  
\begin{tabular}{cc}  
\resizebox{50mm}{!}{\includegraphics{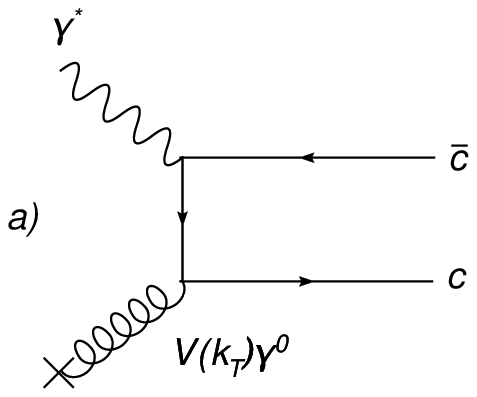}} & 
\resizebox{50mm}{!}{\includegraphics{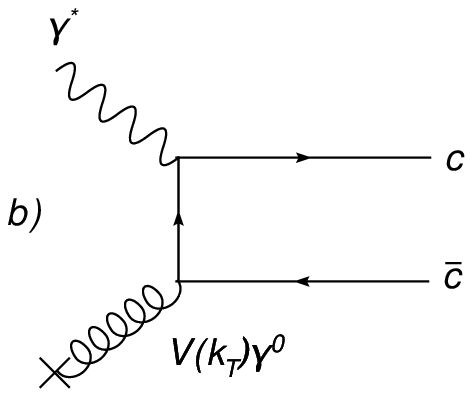}}  
\end{tabular}  
\caption{Boson-Gluon-Fusion graphs. FIG (2a) shows $iM_{1}$, 
FIG (2b) $iM_{2}$.}  
\label{Feynm}  
\end{center}  
\end{figure}

The $\gamma^{*}p$ charm quark production cross section is calculated in the
dipole model (cf. ref. \cite{Kopeliovich:2003cn}).  We start with the
photon-gluon-fusion processes shown in fig.~\ref{Feynm}.  The respective
Feynman graphs are represented by the amplitudes $M_{1}^{\mu}$ and
$M_{2}^{\mu}$\\
  
\begin{eqnarray}  
iM_{1}^{\mu}&=&  
-\sum_{\sigma}\varepsilon_{\mu}(\lambda)\bar u_{\sigma_{q}}(p_{q})  
V(k_{t})\gamma^{0}\frac{u_{\sigma}(q-p_{p_{\bar q}})  
\bar u_{\sigma}(q-p_{p_{\bar q}})}{(q-p_{\bar q})^{2}-m^{2}}ee_{Q}  
\gamma^{\mu}v_{\sigma_{\bar q}}(p_{\bar q}) \nonumber\\  
iM_{2}^{\mu}&=&  
\sum_{\sigma}\varepsilon_{\mu}(\lambda)\bar u_{\sigma_{q}}(p_{q})  
ee_{Q}\gamma^{\mu}\frac{v_{\sigma}(q-p_{q})\bar   
v_{\sigma}(q-p_{q})}{(q-p_{\bar q})^{2}-m^{2}}V(k_{t})\gamma^{0}v_{\sigma_{\bar q}}(p_{\bar q}).  
\label{eqn:amplsum}  
\end{eqnarray}  
  
Here $\varepsilon(\lambda)$ is the polarization vector of the virtual photon  
$\gamma^{*}$ with polarization $\lambda$, $u_{\sigma_{q}}(p_{q})$   
the spinor of the heavy quark with helicity   
$\sigma_{q}$ and momentum $p_{q}$, and $v_{\sigma_{\bar q}}(p_{\bar q})$   
the spinor of the antiquark with helicity $\sigma_{\bar q}$ and momentum $p_{\bar q}$.  
The charge of the heavy quark in units of $e$ is $e_{Q}$.  
The polarization vectors are given in eqs. (\ref{eqn:polvector}).  
The target color field represented by the vertex function $V(k_{t})$ depends on    
the transverse momentum $k_{t}$ injected by the gluon from the target.  
The vertex V  is restricted to its time-component,   
i.e. the formula holds only in the target rest frame.   
It is assumed that the target gluon mediates transverse momentum only,   
longitudinal momentum and energy transfers are negligible.    
In the high-energy approximation the fermion propagators can be treated onshell   
\cite{he}. This approximation is equivalent to neglecting the instantaneous terms   
in light-cone quantization \cite{bj}. We can    
express the energy denominators in terms of LC variables. The coordinate system is   
chosen such that the z-axis points in $\vec q$-direction. With   
the light-cone momentum of the photon $( q^+,0_{\bot}, q^-)=(Q^2/(2 m x_B),0,-m x_B)$  
and  
$\alpha$ as the light-cone momentum fraction of the quark we find    
$(p_{q}^+,p_{q\bot},p_{q}^- )=(\alpha  q^+ ,p_{q\bot},(p_{q\bot}^2+m^2)/(\alpha  q^+))$.  
This allows to simplify the propagators in eq. (24). Equivalent expressions  
hold for the antiquark with $\alpha$ replaced by $1-\alpha$ 
  
\begin{eqnarray}  
\frac{1}{(q-p_{q})^{2}-m^{2}}&=&\frac{-\alpha}{p_{q\bot}^{2}+\epsilon^{2}} \nonumber\\  
\frac{1}{(q-p_{\bar q})^{2}-m^{2}}&=&\frac{\alpha-1}{(p_{q\bot}-k_{t})^{2}+\epsilon^{2} }  
\nonumber\\  
\vphantom{\frac{1}{1}}\epsilon^{2} &=&\alpha(1-\alpha)Q^{2}+m^{2}.  
\label{eqn:edenom}  
\end{eqnarray}

The parameter $\epsilon^{2}$   
is related to the inverse extension squared of the quark-antiquark state.  
  
The $\gamma^{*}p$   
cross section  is proportional to the absolute square of the summed transition 
amplitudes $M_{i}^{\mu}$.  
In the phase space integration we use the integration variable $k_t$ given   
by the equation $k_{t}=p_{q}+p_{\bar q}-q$  
instead of $p_{\bar q}$:  
  
\begin{equation}  
d^{5}\sigma=\sum_{\sigma_{q} \sigma_{\bar q} c_{q} c_{\bar q}\lambda,\lambda'}   
\frac{d\alpha\, d^{2}p^{\bot}_{q}\,d^{2}k_{t}}{8(2\pi)^{5}  
(q^{0})^{2}\alpha(1-\alpha)}\epsilon_{\mu}^{*}(\lambda)\epsilon_{\nu}(\lambda')  
(M_{1}^{\mu}+M_{2}^{\mu})(M_{1}^{*\nu}+M_{2}^{*\nu}.)  
\label{eqn:csform}  
\end{equation}

The polarization-interference terms vanish when integrated over   
the azimuthal angle.  
It is advantageous to transform all expressions into impact parameter space since   
at high energies the quarks propagate along fixed  
impact parameter trajectories.  
We express the vector amplitudes  
$M_{i}^{\mu}(\vec k_t,\vec p_q)$  
by their Fourier transforms $\tilde M_{i}^{\mu}(\vec b,\vec r)$ in eq. (27) 
  
\begin{equation}  
M_{i}^{\mu}(\vec k_t,\vec p_q)=\int d^{2}b d^{2}r\tilde M_{i}^{\mu}(\vec b,\vec r)  
\textrm{e}^{i\vec k_{t}\vec b+i\vec p_{q}^{\bot}\vec r}.  
\end{equation}  
  
The explicit form of the vector amplitudes in coordinate space is:  
  
\begin{eqnarray}  
\tilde M_{1}^{\mu}(\vec b,\vec r)&=&-i2p_{q}^{0}ee_{Q}(1-\alpha)  
\int\frac{d^{2}k_{t}\,d^{2}p_{q}^{\bot}}{(2\pi)^{4}}\,  
\textrm{e}^{-i\vec k_{t}\vec b-i\vec p_{q}^{\bot}\vec r}\,V(k_{t})\nonumber\\  
&&\qquad\qquad\times\qquad\frac{\bar   
u_{\sigma_{q}}(q-p_{\bar q})\gamma^{\mu}  
v_{\sigma_{\bar q}}(p_{\bar q})}{(p_{q\bot}-k_{t})^{2}+\epsilon^{2}}\nonumber\\  
\tilde M_{2}^{\mu}(\vec b,\vec r)&=&i2p_{\bar q}^{0}ee_{Q}  
\alpha\int\frac{d^{2}k_{t}\,d^{2}p_{q}^{\bot}}{(2\pi)^{4}}\,  
\textrm{e}^{-i\vec k_{t}\vec b-i\vec p_{q}^{\bot}\vec r}\,V(k_{t})\nonumber\\  
&&\qquad\qquad\times\qquad\frac{\bar u_{\sigma_{q}}(p_{q})  
\gamma^{\mu}v_{\sigma_{\bar q}}(q-p_{q})}{p_{q\bot}^{2}+\epsilon^{2}}.  
\end{eqnarray}

With the vertex function in impact parameter space  
  
\[ \tilde V(\vec b)=\int\frac{d^{2}k_{t}}{(2\pi)^{2}}\exp(-i\vec k_{t}\vec b)V(\vec k_{t}) \]  
  
and the light-cone wave function:

\begin{equation}  
\Psi^{\mu}(\vec r)=e_{Q}\sqrt{\alpha_{em}}\sqrt{N_{c}}  
\sqrt{\alpha(1-\alpha)}\int\frac{d^{2}p_{q}^{\bot}}  
{(2\pi)^{2}}\,\textrm{e}^{-\vec p_{q}^{\bot}\vec r}\,  
\frac{\bar u_{\sigma_{q}}(p_{q})  
\gamma^{\mu}v_{\sigma_{\bar q}}(q-p_{q})}{p_{q\bot}^{2}+\epsilon^{2}}  
\label{eqn:psialsint}  
\end{equation}

one finds that the  momentum space integrals  factorize  
and yield the convenient expressions:  
  
\begin{equation}  
\begin{array}{rcl}  
\tilde M_{1}^{\mu}(\vec b, \vec r)&=&-i2p_{q}^{0}  
\sqrt{\frac{1-\alpha}{\alpha}}\sqrt{\frac{4\pi}{N_{c}}}  
\tilde V(\vec b+\vec r)\Psi^{\mu}(\vec r)\\\\  
\tilde M_{2}^{\mu}(\vec b, \vec r)&=&i2p_{\bar q}^{0}  
\sqrt{\frac{\alpha}{1-\alpha}}\sqrt{\frac{4\pi}{N_{c}}}\tilde V(\vec b)\Psi^{\mu}(\vec r).  
\label{eqn:amplimp}  
\end{array}  
\end{equation}

The light-cone wave function $\Psi^{\mu}$ describes   
the splitting $\gamma^{*}\rightarrow q\bar q$ and appears  identically in both   
initial state amplitudes.  
The vertex function has the respective arguments $b$ and $b+r$ which are the   
impact parameters of the antiquark in $M_{2}$ and $M_{1}$.    
Using $p_{q}^{0}\approx\alpha q^{0}$, $p_{\bar q}^{0}\approx (1-\alpha)q^{0}$,  
we find for the $\gamma^{*}p$ production cross section of a heavy quark with  
momentum $p_q$

\begin{eqnarray}  
\frac{d\sigma^{L,T}}{d^{2}p_{q}^{\bot }}&=&  
\int{\frac{1}{(4\pi)^{2}}\int d^{2}b d^{2}r_{1}\,d^{2}r_{2}  
d\alpha\,\textrm{e}^{i\vec p_{q}^{\bot}(\vec r1-\vec r2)}}\\  
&&\times\frac{1}{N_{c}}\sum_{c_{q}c_{\bar q}}\Psi^{L,T}(\vec r_{1})\Psi^{*L,T}(\vec r_{2})  
[\tilde V(\vec b)-\tilde V(\vec b+\vec r_{1})]  
[\tilde V^{\dag}(\vec b-\tilde V^{\dag}(\vec b+\vec r_{2})]. \nonumber  
\label{eqn:halbfertig}  
\end{eqnarray}   
  
Due to the integration over the antiquark momentum which is equivalent to the $k_t$  
integration,   
the impact parameters in $(\tilde M_{1}^{\mu}+\tilde M_{2}^{\mu})$  
and $(\tilde M_{1}^{*\nu}+\tilde M_{2}^{*\nu})$ coincide. Since the momentum of the  
quark $p_q$ remains unintegrated,  the exponentials of the Fourier transforms survive.  
The light-cone wave functions simplify after summation  over spins and polarizations  
  
\begin{equation}  
\begin{array}{rcl}  
\Psi^{L}(\vec r_{1})\Psi^{*L}(\vec r_{2})&=&  
\sum_{\sigma_{q}\sigma_{\bar q}}\varepsilon_{\mu}(\lambda=0)  
\varepsilon^{*}_{\nu}(\lambda=0)\Psi^{\mu}(\vec r_{1})\Psi^{*\nu}(\vec r_{2})\\\\  
\Psi^{T}(\vec r_{1})\Psi^{*T}(\vec r_{2})&=&  
\frac{1}{2}\sum_{\lambda=\pm\frac{1}{2}}\sum_{\sigma_{q}\sigma_{\bar q}}  
\varepsilon_{\mu}(\lambda)\varepsilon^{*}_{\nu}(\lambda)\Psi^{\mu}  
(\vec r_{1})\Psi^{*\nu}(\vec r_{2}).  
\label{eqn:lcwfLT}  
\end{array}  
\end{equation}  
  
Combining the gluon vertex operators  $V$ appropriately we   
can construct dipole cross sections 
  
\begin{eqnarray}  
\!\!\!\!\!\!\!\!&&\lefteqn{\int d^{2}b [\tilde V(\vec b)-\tilde V(\vec b+\vec r_{1})]  
[\tilde V^{\dag}(\vec b)-\tilde V^{\dag}(\vec b+\vec r_{2})]=}\nonumber\\  
\!\!\!\!\!\!\!\!&&  
\frac{1}{ 2}  
\int d^{2}b  
\left[\mid \tilde V(\vec b)-\tilde V(\vec b+\vec r_1)\mid^{2}  
+  
\mid \tilde V(\vec b)-\tilde V(\vec b+\vec r_2)\mid^{2}  
-  
\mid \tilde V(\vec b+\vec r_2)-\tilde V(\vec b+\vec r_1)\mid^{2}\right].  
\end{eqnarray}  
  
The corresponding absolute squares can be grouped into dipole cross  
sections employing the definition:  
  
\begin{equation}  
\sigma_{q\bar q}(r)=  
\frac{1}{N_{c}}\sum_{c_{q}c_{\bar q}}  
\int d^{2}b\mid \tilde V(\vec b)-\tilde V(\vec b+\vec r)\mid^{2} . 
\label{eqn:dcsv}  
\end{equation}  
  
One finds the following final form for the differential $\gamma^{*}$p cross section  
for charm quark production, i.e. $q=c$:

\begin{eqnarray}  
\frac{d\sigma^{L,T}(\gamma^{*}p\rightarrow(c X))}  
{d^{2}p_{c}^{\bot}}&=&\frac{1}{(2\pi)^{2}}\int d^{2}r_{1}\,d^{2}r_{2}\,d\alpha\,  
\textrm{e}^{i\vec p_{c}^{\bot}(\vec r_{1}-\vec r2)}\,  
\Psi^{L,T}(\vec r_{1})\Psi^{*L,T}(\vec r_{2})\nonumber\\  
&\times&\frac{1}{2}\lbrace\sigma_{c\bar c}(\vec r_{1})+  
\sigma_{c\bar c}(\vec r_{2})-\sigma_{c\bar c}(\vec r_{1}-\vec r_{2})\rbrace  .
\label{eqn:gammap}  
\end{eqnarray}

A similar expression has been derived in \cite{kst1} to describe the  
transverse momenta of Drell-Yan pairs. In the differential cross  
section light-cone wave functions enter with   
Bessel functions $K_0$ and $K_1$ depending on  
$\varepsilon=\sqrt{m^2+\alpha(1-\alpha)Q^2}$ and $r_1$ or $r_2$:  
\begin{eqnarray}   
\Psi^{L}(\alpha,r_{1})\Psi^{*L}(\alpha,r_{2})&=  
&\frac{2N_{c}\alpha_{em}e_{Q}^{2}}{(2\pi)^{2}}\,4Q^{2}  
\alpha^{2}(1-\alpha)^{2}K_{0}(\epsilon r_{1})K_{0}(\epsilon r_{2})\nonumber\\  
\Psi^{T}(\alpha,r_{1})\Psi^{*T}(\alpha,r_{2})&=  
&\frac{6e_{Q}^{2}\alpha_{em}}{(2\pi)^{2}}\lbrace m_{Q}^{2}K_{0}  
(\epsilon\vec r_{1})K_{0}(\epsilon\vec r_{2})\nonumber\\  
&+&\epsilon^{2}[\alpha^{2}+(1-\alpha)^{2}]  
\frac{\vec r_{1}\vec r_{2}}{r_{1}r_{2}}K_{1}(\epsilon\vec r_{1})
K_{1}(\epsilon\vec r_{2})\rbrace.  
\end{eqnarray}  
  
If one integrated out the transverse quark momentum in equation  
(\ref{eqn:gammap}), one would find a delta function in $(r_{1}-r_{2})$  
and recover the total cross section of charm production. The differential cross  
section does not have such an intuitive interpretation as the total DIS cross  
section. As shown in the derivation a physical  
quark-antiquark fluctuation is not a prerequisite for the appearance  
of the dipole cross section in the inclusive cross section. Instead,  
interference terms in the square of the transition amplitude produce  
dipole cross sections when a  quark in the amplitude and a quark in  
the complex conjugate amplitude enter with  different impact parameters.   
This becomes very obvious in the Drell-Yan process, where the dipole formula   
for the transverse momentum distribution of lepton pairs looks very   
similar to (\ref{eqn:gammap}) although there is not necessarily any   
dipole fluctuation involved in the  Feynman graphs \cite{hir}.  
  
We gave the inclusive DIS cross section in terms of the  
cross section of a quark-antiquark pair scattering  off a  
proton. For the  calculations in this paper we use a fit to the dipole cross  
section and its energy dependence. For an  explicit derivation of the dipole  
cross section from QCD we refer to the loop-loop correlation model, see  
Ref. \cite{Shoshi:2002in}.   
In the following we use the GBW-saturation model by Golec-Biernat and  
W\"usthoff \cite{gbw} with a dipole-nucleon cross section of the form:  
\begin{equation}  
\sigma_{q \bar q}(x,r)=\sigma_{0}\left[1-\exp\left(-\frac{r^{2}}{4R_{0}^{2}(x)}\right)\right]  
\label{eqn:gbw}  
\end{equation}  
where  
\begin{equation}  
R_{0}=\frac{1}{Q_{0}}\left( \frac{x}{x_{0}}\right)^{\frac{\lambda}{2}}.  
\label{eqn:r0}  
\end{equation}  
This cross section is not a well-defined quantity for  
$Q^2=0$. Therefore, $x$ has been modified  
\begin{equation}  
x\rightarrow x\left( 1+\frac{4m_{Q}^{2}}{Q^{2}} \right)  
\end{equation}  
in order to describe the transition to the photoproduction region. The  
distance between the quark and antiquark is $r$. The inverse momentum  $1/Q_{0}=0.2 fm$   
is a perturbative distance. At $x=x_0=0.41*10^{-4}$ twice this distance equals   
the distance where the cross section  
changes from the perturbative $r^2$ behavior into a constant behavior.   
The parameters $\lambda, \sigma_{0}$  
and $x_{0}$ are given in Table (\ref{tab:gbwpara}) \cite{gbw}. The fit  
is performed within a Bjorken interval  $4*10^{-4}\leq x_{B}\leq 0.01$  
and  virtuality range of $0.1\textrm{GeV}^{2}\leq Q^{2}\leq 400\textrm{GeV}^{2}$.   
\begin{table}[htbp]  
\begin{center}  
\begin{tabular}{l||r|r|r|c}  
 &  $\sigma_0$ (mb) & $~~~~\lambda~~~~$ & ~~$x_0$~~~~~~~~& ~~~~$1/Q_0(fm)$~~~\\  
\hline   
$\sigma_{q\bar q}$ with charm   & ~~29.12~~ & ~~0.277~~ &~~ $0.41\cdot10^{-4}$~~&~~0.2    \\  
\end{tabular}  
\caption{Fit parameters in  the GBW saturation model \cite{gbw}.   
The charm quark mass is fixed to $m=1.5$GeV.}  
\label{tab:gbwpara}  
\end{center}  
\end{table}  
  
The saturation model contains four fit parameters. This is a small  
number compared to the number of parameters in parton distribution  
functions. The authors have performed a fit to HERA data and reached a  
satisfactory $\chi^{2}$ in Ref.~\cite{gbw} containing charm in the sum over flavors.  
  
The model describes two different saturation effects. The GBW cross  
section saturates for  quark-antiquark pairs with large $r$  
separation. Arbitrarily large $q\bar q$-separations are not  
physical due to hadronization, but this does not matter, since 
they occur in the $q\bar q$ light-cone wave function with very low probability.   
The other saturation effect is implemented via  the $x$-dependence of  
$R_{0}\sim 1/Q_{s}$, where $Q_s$ acts as a saturation scale. The  
$\gamma^{*}p$ cross section rises earlier to $\sigma_0$  for decreasing   
$x$. One expects this increase to be slowed down or stopped at even  
lower $x$ when the gluon-gluon-recombination cross section 
becomes sizeable \cite{dis}. The color-glass model predicts
a new QCD domain which is nonperturbative but not dominated by confinement effects.  
Note, both saturation effects are not to be mixed up with   
the saturation of the gluon density in  impact parameter space 
discussed in Ref. \cite{Shoshi:2002in}.  
  
For our calculations we can expand the cross section to first order in $r^{2}$  
\begin{equation}  
\sigma_{q \bar q}(x,r)\simeq\sigma_{0}\frac{r^{2}}{4R_{0}^{2}(x)}=:\tilde\sigma_{0}r^{2},  
\label{eqn:r^2}  
\end{equation}  
where $R_{0}$ is given in (\ref{eqn:r0}).  
In the following we refer to this use of the color dipole cross section as  
 $r^{2}$-approximation. For $D$-meson production with a charm  mass  $m=1.5$ GeV, we do  
not expect that the size of the semi-inclusive  cross section is very much influenced   
by large dipole separations. The dipoles contributing have a size of about 0.2 fm.   
  
\section{Numerical calculation}  
\label{sec:num}

With the analytical formulae and the above dipole cross section we can numerically
calculate the double differential charmed quark production cross section
$d\sigma^{T,L}/d^{2}p_{c}^{\bot}$.  We reduce the fourfold Fourier integral over $\vec
r_{1}$, $\vec r_{2}$ which occurs in the $\gamma^{*}p$ sub cross section
(\ref{eqn:gammap}) to a one-dimensional integral over the dipole separation $r$. We
find integrals with modified Bessel functions of the second kind $K_{0,1}$ and Bessel
functions of the first kind $J_{0,1}$
  
\begin{eqnarray}  
\frac{d\sigma(\gamma^{*}p\rightarrow c\,X)}{d^{2}p_{c}^{\bot}}  
&=&\frac{6e_{Q}^{2}\alpha_{em}}{(2\pi)^{2}}\int d\alpha\nonumber\\  
&\times&\left\lbrace\left[\vphantom{\frac{1}{1}}m_{c}^{2}+
4Q^{2}\alpha^{2}(1-\alpha)^{2}\right]  
\right.\left[\frac{I_{1}}{p_{c}^{\bot 2}+\epsilon^{2}}-\frac{I_{2}}{4\epsilon}\right]
\nonumber\\  
&+&\left[\vphantom{\frac{1}{1}}\alpha^{2}+(1-\alpha)^{2}\right]\left.  
\left[\frac{p_{c}^{\bot}\epsilon I_{3}}{p_{c}^{\bot 2}+  
\epsilon^{2}}-\frac{I_{1}}{2}+\frac{\epsilon I_{2}}{4}\right]\right\rbrace  
\label{eqn:epI}  
\end{eqnarray}  
with   
\begin{eqnarray}  
I_{1}&=&\int dr\,r\,J_{0}(p_{c}^{\bot}r)\,K_{0}(\epsilon r)\,\sigma_{q\bar q}(r)\nonumber\\  
I_{2}&=&\int dr\, r^{2}\, J_{0}(p_{c}^{\bot}r)\,K_{1}(\epsilon r)\,\sigma_{q\bar q}(r)\nonumber\\  
I_{3}&=&\int dr\, r\, J_{1}(p_{c}^{\bot}r)\,K_{1}(\epsilon r)\,\sigma_{q\bar q}(r)\ .  
\label{eqn:I1-I3}  
\end{eqnarray}  
  
\begin{figure}[htbp]  
\begin{center}  
\resizebox{10cm}{!}{\includegraphics{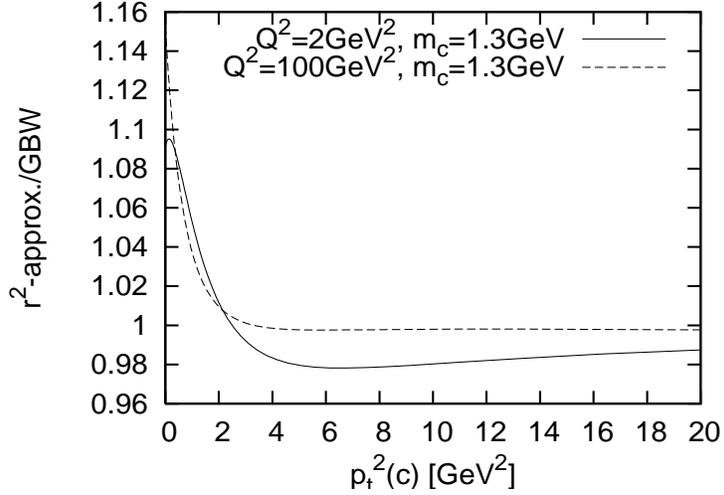}}  
\caption{The ratio of the cross section $d\sigma(\gamma^{*}p\rightarrow  
  c\,X)/dp_{c}^{\bot 2}$ in $r^{2}$-approximation over the same cross  
  section calculated with the full GBW cross section $\sigma(r)$ is shown     
as a function of the transverse momentum of the charm quark $p_{c}^{\bot 2}$.}  
\label{fig:fullratio}  
\end{center}  
\end{figure}

The small $r^{2}$-approximation (\ref{eqn:r^2}) of  the GBW dipole cross section  
allows to perform the $r$-integrations $I_{1}$-$I_{3}$ analytically 
  
\begin{eqnarray}  
\left.I_{1}\right\vert_{r^{2}-approx.}  
&=&\frac{\sigma_{0}}{4R_{0}^{2}}\,\frac{4\,  
(\epsilon^{2}-p_{c}^{\bot 2})}{(p_{c}^{\bot 2}+\epsilon^{2})^{3}}\nonumber\\  
  \left.I_{2}\right\vert_{r^{2}-approx.}  
&=&\frac{\sigma_{0}}{4R_{0}^{2}}\,\frac{16\,\epsilon\,  
(\epsilon^{2}-2p_{c}^{\bot 2})}{(p^{\bot 2}_{c}+\epsilon^{2})^{4}}\nonumber\\  
  \left.I_{3}\right\vert_{r^{2}-approx.}  
&=&\frac{\sigma_{0}}{4R_{0}^{2}}\,\frac{8\,p^{\bot}_{c}\,  
\epsilon}{(p_{c}^{\bot 2}+\epsilon^{2})^{3}}\ .  
\label{eqn:i1-3ana}  
\end{eqnarray}  
The integrals  require an extension parameter $\epsilon$ greater than zero  
to be finite.   
For heavy-quark production with $\epsilon \geq m $ this is no problem.  
We have checked how much the approximate $r^{2}$-calculation and the  
full GBW dipole cross section differ for the sum of transversely and  
longitudinally polarized photons. In Fig.~\ref{fig:fullratio} we  
show the $\gamma^{*}p\rightarrow c\,X$ cross section in  
$r^{2}$-approximation over the same cross section calculated with the  
full GBW parametrization. Deviations appear mainly at low transverse  
momenta of less than about $1\ \textrm{GeV}^{2}$. 
We find good agreement for higher $Q^{2}$,  
when the average dipole size is small.   
Above $p_{c}^{\bot 2}> 2\ \textrm{GeV}^2$ we find a  
very small deviation of maximally 2\%, which is sufficient for our calculations.  
  
\begin{figure}  
\begin{center}  
\resizebox{9.5cm}{!}{\includegraphics{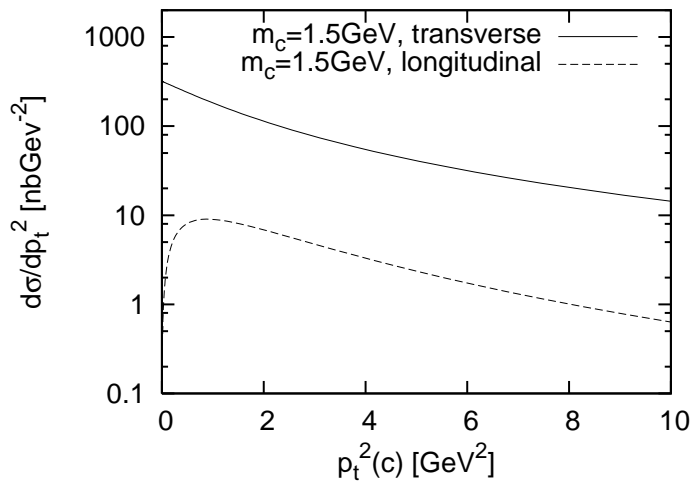}}  
\caption{The charm quark   
production cross section $\gamma^{*}p\rightarrow c X$ induced by transverse and  
longitudinal photons  for $Q^{2}=2\ \textrm{GeV}^{2}$ and $W^{2}=200\ \textrm{GeV}^{2}$.    
The transverse photon cross section is dominating over the whole  
transverse momentum region.}  
\label{fig:mvgllt}  
\end{center}  
\end{figure}

Next, we study the relative contributions of longitudinally and  
transversely polarized virtual photons in Fig.~\ref{fig:mvgllt}.  
The transverse cross section is clearly dominating. The relative  
contribution of the longitudinal cross section grows when the virtuality  
of the photon changes from  
$Q^{2}=2\ \textrm{GeV}^{2}$ to $100\ \textrm{GeV}^{2}$.  
This behavior is similar to vector meson production.  

To calculate D-meson cross sections we must let the charm quark fragment.    
The fragmentation function $D_{h}^{Q}(z^{*})$ gives the probability   
that the  
original charm quark with a momentum $P$    
fragments into a D-meson with momentum fraction $z^* P$.  
All our momentum fractions in section 3 refer to the photon momentum $q$.  
Since we start with a charm quark of momentum $P=\alpha q$ and end up with a  
D-meson with momentum $z q$ the momentum fractions multiply:  
  
\begin{equation}  
  z=z^{*}\alpha.  
\end{equation}

To calculate D-meson production we convolute the charm quark  
production cross section $d\sigma^{L,T}(\gamma^{*}p\rightarrow cX)$  
with the nonperturbative fragmentation function $D(z^{*})$:  
$d\sigma^{L,T}(\gamma^{*}p\rightarrow DX)$  
\begin{eqnarray}  
  \frac{d\sigma^{L,T}}{dz\,d^{2}p_{D}^{\bot}}  
(\gamma^{*}p\rightarrow DX)&=&\int\frac{dp_{c}^{\bot}\,d\alpha}{\alpha}\,  
\frac{d\sigma^{L,T}}{d^{2}p_{c}^{\bot}\,d\alpha}(\gamma^{*}p\rightarrow cX)\\  
  &\times&D^{c}_{D}\left(\frac{z}{\alpha}\right)\,\delta\left(p^{\bot}_{D}-\frac{z}  
{\alpha}\,p^{\bot}_{c}\right)\ .  
\label{eqn:conv. mit z}  
\end{eqnarray}  
  
We use the Peterson fragmentation function \cite{pe} which has the form  
  
\begin{equation}  
D_{Q}^{h}(z^{*})=\frac{n(h)}{z^{*}[1-\frac{1}{z^{*}}-\frac{\epsilon_{Q}}{1-z^{*}}]^{2}}\ .  
\label{eqn:pe}  
\end{equation}  
  
This fragmentation function  has been successful in describing experimental data   
of heavy flavored hadrons and can be easily understood in the following way.   
The transition amplitude is  
proportional to the inverse energy difference $\Delta E$ between the  
initial heavy quark $Q$ and the final meson $h$ plus light quark state  
$q$  
\begin{eqnarray}  
\Delta E &=& E_{Q}-E_{h}-E_{q}\nonumber\\  
&=& \sqrt{m_{Q}^{2}+P^{2}}-\sqrt{m_{h}^{2}+z^{2}P^{2}}-\sqrt{m_{q}^{2}-(1-z)^{2}P^{2}}\nonumber\\  
&\approx& \frac{m_{Q}^{2}}{2P}\left(1-\frac{1}{z^{*}}-\frac{e_{Q}}{1-z^{*}}\right)\ ,  
\label{eqn:deltaE}  
\end{eqnarray}  
where $m_{h}=m_{Q}$ can be used for simplicity. The fragmentation function is
given by the square of the transition amplitude normalized correctly. The Peterson   
parameter $e_{Q}=m^{2}_{q}/m^{2}_{Q}$ parametrizes the hardness of the  
fragmentation and is fitted to data. The larger the quark mass, the smaller is $\epsilon_{Q}$   
and the harder  is the fragmentation, i.e. $D_{Q}^{h}(z^{*})$ is peaked more towards $z^{*}=1$.  
The z-integrals of the Peterson fragmentation functions  $D_{Q}^{h}$ are normalized to  
the branching ratios $ f_{Q}^{h}$ which summed over all hadron species add up to unity  
  
\begin{equation}  
\sum_{h}f_{Q}^{h}=\sum_{h}\int D_{Q}^{h}(z^{*})dz^{*}=1.  
\end{equation}  
  
In general we are using the branching ratios given in Ref.  \cite{h1}   
which refer to the specific kinematical domain of the H1-experiment.
 The world averages  evaluated by the Particle Data Group \cite{Yao:2006px}   
are also listed   
in Table (\ref{tab:br}) for comparison. The large isospin violating branching ratio  
for the $D^0$-meson   
compared with the $D^+$-meson branching ratio comes from the accidental nearness  
of the D$^{*}$-resonances to the sum of the masses of the D-meson and a pion. Namely   
both the D$^{*+}$ and $D^{*0}$ can decay into $D^0$ with large branching fractions,  
but the  D$^{*+}$ can only decay to the $D^+$ with a small branching.  
The Peterson parameter for the charm fragmentation function is  
$\epsilon_{c}=0.05$ in  the leading-logarithmic approximation  
(LLA) \cite{pdg}.   
Although the branching ratios into D-mesons and $D^*$-mesons are approximately equal,  
it is much easier to measure $D^*$- mesons than D-mesons at HERA, since $D^*$-mesons   
have a good signature due to their $D^*\rightarrow D\pi \rightarrow K \pi\pi$ decay chain.  
The fits are performed with data from `clean' $e^{+}e^{-}$ reactions.  
We use these functions in DIS assuming independence from the underlying  
production process of the heavy quark.

\begin{table}[htbp]  
\begin{tabular}{c|cccc}  
$c\rightarrow$ & $D^{0}$ & $D^{+}$ & $D^{*0}$ & $D^{*+}$\\  
\hline  
PDG & $0.565\pm 0.032$ & $0.246\pm 0.020$ & $0.213\pm 0.024$ & $0.224\pm 0.028$\\  
H1 & $0.658\pm 0.054$ & $0.202\pm 0.020$ &-- & $0.263\pm0.019$\\  
\end{tabular}  
\caption{Branching ratios $c\rightarrow D$ for several $D$-mesons.   
Reference values from Particle Data Group PDG \cite{pdg} and H1 \cite{h1}.   
The H1 data is fitted without theoretical constraints due to the experimentally   
covered kinematical region.}  
\label{tab:br}  
\end{table}

\begin{figure}[htbp]  
\begin{center}  
  \resizebox{12cm}{!}{\includegraphics{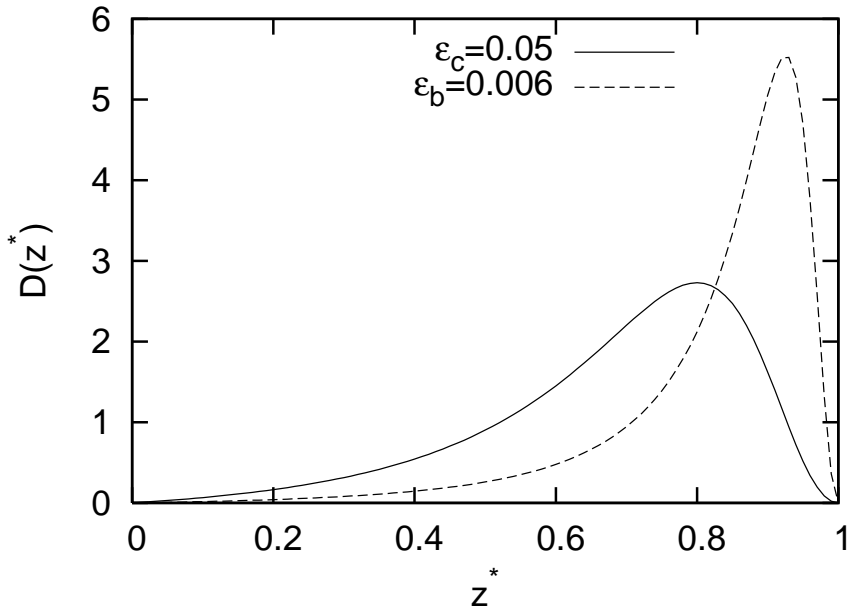}}  
  \caption{The Peterson fragmentation function with $\epsilon_{c}=0.05$  
   for charm quarks and  $\epsilon_{b}=0.006$ for bottom quarks  
    in the leading log approximation. The bottom quark fragmentation is harder since the  
    bottom quark mass is about 3.5 times higher than the charm quark  
    mass. Both curves are normalized to unity.}  
\label{fig:peterson}  
\end{center}  
\end{figure}

The fragmentation process of bottom quarks requires a harder  
fragmentation function since the b-quark mass is 4-4.4 GeV, i.e. about 3.5  
times larger than the charm quark mass. The reference value   
 $\epsilon_{b}=0.006$ for the bottom quark is therefore much smaller in the  
leading logarithmic approximation. The attachment of a  
light quark degrades the bottom quark momentum less than the charm  
quark momentum. The LLA-Peterson fragmentation functions are plotted for  
both flavors in Fig.~\ref{fig:peterson}.  

\begin{figure}[htbp]  
\begin{center}  
\resizebox{12cm}{!}{\includegraphics{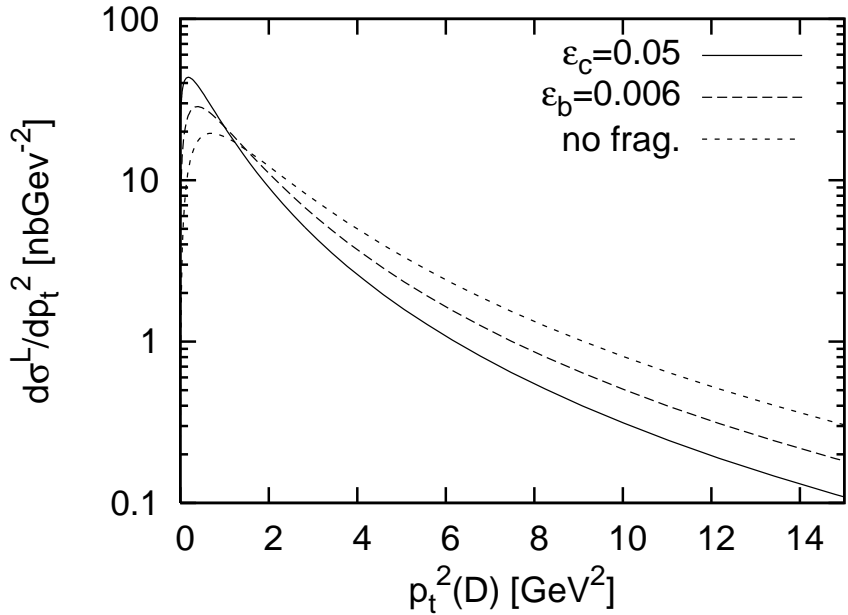}}  
\caption{The longitudinal heavy meson production cross section $\gamma^{*}p\rightarrow h(Q) X$   
induced by longitudinal photons as a function of the transverse momentum of the meson.  
The two $\epsilon$ parameters correspond to charm ($\epsilon_c=0.05$) and   
bottom quark ($\epsilon_b=0.006$) fragmentation.    
For comparison we also show the transverse quark   
production cross section without fragmentation.}  
\label{fig:Dvgl}  
\end{center}  
\end{figure}

When a high-momentum charm  quark in the photon fragments into a   
charmed meson without additional   
transverse momentum, the scaling variable $z^*$   
gives the transverse and longitudinal fractions $p_c^{\bot}/p_D^{\bot}=p_c^L/p_D^L=1/z^*$  
at the same time.   
One finds:  
\begin{displaymath}  
<p_{c}^{\bot}/p_{D}^{\bot}>=\sum_{h}   
\int _{0}^{1}\frac{dz^{*}}{z^{*}}D^h_c(z^*)  
\end{displaymath}  
a mean ratio of transverse momenta $<p_c^\bot/p_{D}^{\bot}>=1.68$ and $<p_b^\bot/p_B^\bot>=1.29$
for the bottom sector, respectively. This makes the differential cross sections   
of the heavy quark harder than the meson cross sections by the corresponding ratios.   
In order to see how the fragmentation  
acts we plot the longitudinal $\gamma^{*}p$ cross section  
$d\sigma^L/d^2p^2_t$ for the two different values of $\epsilon_{Q}$ as a  
function of the transverse momentum in Fig.~\ref{fig:Dvgl}. One  
sees how harder fragmentation functions shift the curve to higher  
transverse momenta.  

\section{Comparison with experimental data}  
We evaluate numerically the $ep$ cross section (\ref{eqn:epges}) for  
the H1 experiments \cite{h1,h1alt} at HERA taking into account the  
experimental cuts. In order to specify these cuts we use   
two different coordinate systems, the K-system and the K$^*$-system:   
The $z$-axis of the K-system  
points in the incoming electron beam direction. The  
$z$-axis of the  K$^*$-system  points in the photon direction.  
The experimental results of the measurements by  
the H1 collaboration are given in the K-system, i.e.   
with respect to the electron beam.  
The photon virtuality varies   
$2<Q^{2}<100\,\textrm{GeV}^{2}$, the inelasticity $0.05<y<0.7$ and  
pseudorapidity $\vert \eta(D) \vert\leq 1.5$.  Further the minimum  
transverse momentum $p_{D\,min}^{\bot}$ is $2.5$~GeV for the center-of-mass energy $\sqrt{s}=$319~GeV \cite{h1} and $p_{D\,min}^{\bot}=1.5$~GeV  
for $\sqrt{s}=300$~GeV \cite{h1alt}.  The theoretical dipole  
cross sections we use include well this $Q^{2}$-range.  Our  
calculation, however, extrapolates somewhat the Bjorken-$x$ range used  
in the  
GBW dipole cross section. The GBW range includes the   
$x_B$-interval $[10^{-4},10^{-2}]$,   
the experimental data spread over the larger interval $[10^{-5},0.02]$.   
Therefore the theoretical results for the smallest $x_B$ values should  
be regarded with caution.   
We convert the differential cross section for electron-proton  
scattering (\ref{eqn:epges}) to the experimentally given variables.  
For $x$,$y$,$Q^{2}$ we use the relation  
$s=m_{e}^{2}+M_{P}^{2}+Q^{2}/(xy)$, i.e. $Q^{2}=xys$ in the high-energy approximation.  The rapidity is approximated by the  
pseudo-rapidity, $y_{rap.}\simeq\eta$. In the target rest frame   
(TRF) where $p^{0}_D=m_{T}\cosh\eta$  and $\nu=ys/(2M)$, the z-fraction of  
the D-meson is calculated as follows  
  
\begin{equation}  
z=\frac{P\cdot p_{D}}{P\cdot q}\buildrel{\scriptscriptstyle\rm TRF}  
\over=\frac{p_{D}^{0}}{\nu}=\frac{2M}{ys}m_{T}\cosh{\eta}\ .  
\label{eqn:etaz}  
\end{equation}

\begin{figure}  
\begin{center}  
\resizebox{10cm}{!}{\includegraphics{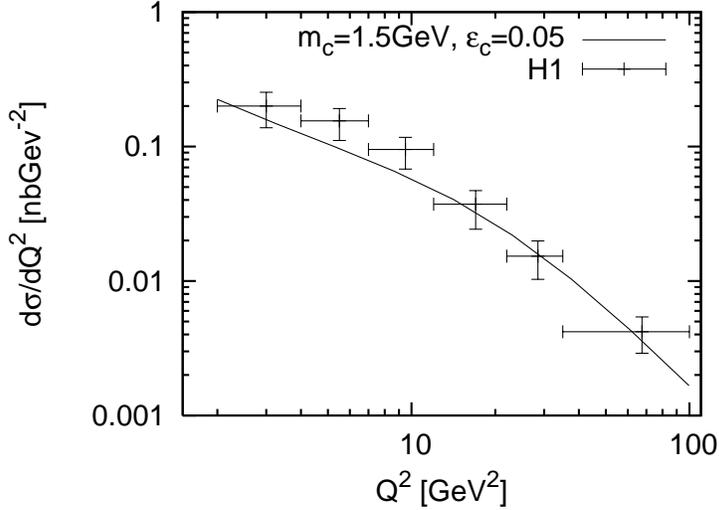}}  
\caption{Theoretical cross section $d\sigma(ep\rightarrow  
  D^{+}X)/dQ^{2}$ with branching ratio $f_{c}^{D^{+}}=0.202$ plotted  
  against experimental data from reference \cite{h1}. Note, the experimental  
  lower cut for the transverse momentum is $p_{D,min}^{\bot}=2.5$~GeV.}  
\label{fig:ecvglQ2}  
\end{center}  
\end{figure}

We evaluate the differential $D^+$-cross sections with respect to  
the photon virtuality $d\sigma/dQ^{2}$ and  
the rapidity of the $D^+$-meson $d\sigma/d\eta$.   
For both cross sections we integrate  
over the $p_{D}^{\bot}$ range inside the experimental cuts.  
In Fig.~\ref{fig:ecvglQ2} we show $d\sigma(ep\rightarrow  
D^{+}X)/dQ^{2}$ for $\epsilon_{c}=0.05$ and $m_{c}=1.5$~GeV and compare  
with the H1 data \cite{h1}. The $Q^2$-decrease is determined by the  
integrals (c.f. eq. (42)) over the  Bessel functions  
$K_{0,1}(\epsilon r)$  
depending on the inverse extension parameter $\epsilon^2=\alpha  
(1-\alpha)Q^2+m^2$ .   
With increasing  
photon virtuality $Q^2$, the Bessel functions damp the size of  
the cross section.  
  
\begin{figure}   
\begin{center}  
\resizebox{10cm}{!}{\includegraphics{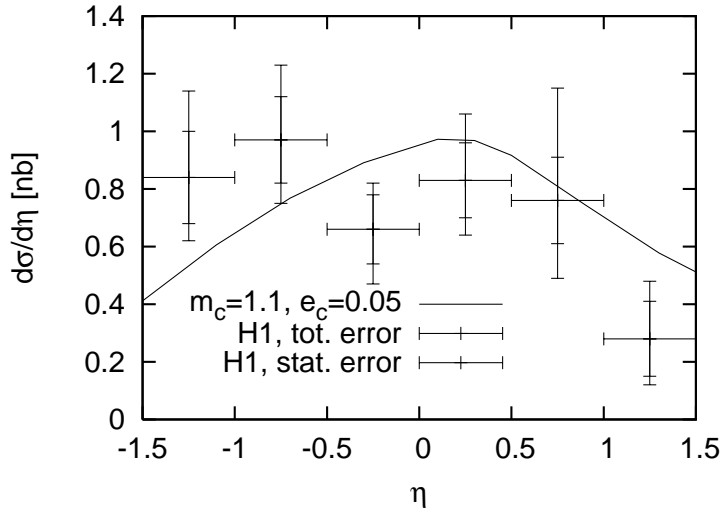}}  
\caption{Theoretical cross section $d\sigma(ep\rightarrow  
  D^{+}X)/d\eta$ with branching ratio $f_{c}^{D^{+}}=0.202$ plotted  
  against experimental data from Ref. \cite{h1}. }  
\label{fig:detavgl}  
\end{center}  
\end{figure}

In Fig.~\ref{fig:detavgl} we plot the $d\sigma(ep\rightarrow  
D^{+}X)/d\eta$ cross section against the H1 data \cite{h1}. For the  
total cross section, higher quark masses would result in a lower $\eta$  
distribution. This holds for the whole rapidity range, even more so   
at mid rapidity. The whole curve becomes higher for a harder  
fragmentation. 
This is expected since the differential cross section falls off for high rapidities 
of either sign. Better   
data are announced \cite{Kruger} to come out for D*-mesons. In order to obtain  
theoretical  predictions for other D-mesons one just has to multiply  
the $D^+$ cross section with the appropriate ratio of branching ratios  
(c.f. Table II). For example by multiplying the $D^+$ cross section with  
2.6 one obtains the combined $D^{*+},D^{*-}$ cross sections.  
  
\begin{figure}  
\begin{center}  
\resizebox{10cm}{!}
{\includegraphics{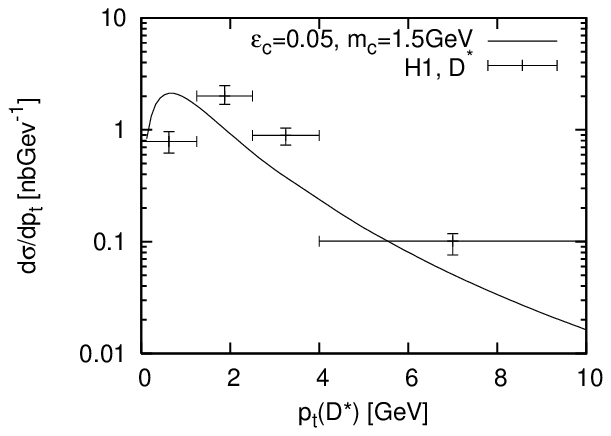}}  
\caption{Theoretical cross section $d\sigma(ep\rightarrow  
  D^{*+}X)/dp_{D}^{\bot}$ with branching ratio $f_{c}^{D^{*+}}=0.263$  
  plotted against experimental data from reference \cite{h1alt}.}  
\label{fig:dpt}  
\end{center}  
\end{figure}

Figure \ref{fig:dpt} shows the theoretical $d\sigma(ep\rightarrow
D^{*}X)/dp_{D}^{\bot}$ cross section against the H1 data \cite{h1alt}. Our theoretical
cross section reproduces the data adequately. It tends, however, to undershoot the
cross section at large $p_T$. This would mean that the gluons in the proton have too
small transverse momenta or the unintegrated gluon density $f_G(x,\vec k_\perp)$
underestimates large transverse momenta. Indeed, the GBW parametrization for the dipole 
cross section leads to a Gaussian $k_T$-dependence for the unintegrated gluon density,
while a power dependence is expected in QCD.

We have also investigated the dependence on the hardness of the fragmentation process.
Our calculation is not very sensitive to changes of the Peterson parameter
$\epsilon_{c}$. A harder fragmentation lifts the differential cross section for
transverse momenta above $\sim 1.5$~GeV and reduces it for lower ones.
  
\section{Summary and outlook}  
  
In this paper the color dipole model has been applied to the production of heavy
flavored mesons in semi-inclusive deep inelastic scattering. In particular, transverse
momentum distributions are a more valuable source of information about the production
and fragmentation processes than total charm production. In the target rest frame, the
virtual photon splits into a heavy quark-antiquark pair which subsequently scatters off
the target. A color dipole is the lowest Fock state of the virtual photon. Instead of
considering higher Fock states containing gluons, which become more important for lower
$x$, these contributions are absorbed into the $x$-dependence of the dipole fit.  The
dipole formulation is in a mixed representation, where the longitudinal momentum is
treated in momentum space and the transverse position in coordinate space. For high
energies, the quark-antiquark pairs with fixed separation are interaction eigenstates.
This formulation is very convenient for treating multiple scattering effects. With the
help of the framework developed in this paper one can also calculate nuclear effects on
heavy quark production by replacing the dipole-proton cross section by the
dipole-nucleus cross section in Glauber approximation.  For electroproduction on the
proton we find a differential cross section in $Q^{2}$ which agrees well with H1 data.
The large phase space coverage helps to collect sufficient statistics in the
experiment. On the other side it presents a real challenge for the theoretical
parametrization of the dipole cross section. In this respect the parametrization of
Golec-Biernat and W\"usthoff's has a clear advantage compared with other approaches due
to its wide kinematical applicability.  It would be interesting to see the effects of
the improved GBW dipole cross section \cite{Bartels:2002cj} on the calculations.
Inclusion of these corrections into the dipole description of direct photon production
\cite{Kopeliovich:2007yv} 
reduces the cross section at high-$p_T$. Nevertheless, our results suggest 
that the
dipole cross section is not the main source of deviations.
  
The fragmentation function has a stronger influence on the shape of the transverse
momentum distribution. We get good fits with a rather hard fragmentation function.
Since there is no previous work which combines hadronization with the dipole approach,
it is worthwhile to investigate how to treat heavy quark production and fragmentation
consistently.  Our fitted dipole cross section contains higher twist effects, i.e. it
includes many soft gluons in the partonic cross section. Therefore we propose to use a
fragmentation function at a low scale without evolution. We do not consider evolution
which may become important at $p_{D}^{\bot}\gg m_{c}$.
  
Intrinsic transverse momenta of the target gluons are encoded into the dipole cross
section. A primordial transverse momentum of the projectile gluon is often introduced
in parton model, even in NLO calculations to harden the spectrum. Usually the 
primordial momentum substantially
exceeds the typical hadronic scales. This is not necessary in the color dipole model,
which generates the intrinsic gluon transverse momenta automatically (e.g. see in 
\cite{Shoshi:2002fq}).

Similar calculation can be applied to hadron induced production of charm \cite{kt}.
While the total cross section of charm production in $pp$ collisions is well explained
\cite{Kopeliovich:2003cn}, it remains to be seen how well $p_T$ distribution can be
explained in the same framework.  Once it is possible to reduce the errors arising from
hadronization, one may obtain more detailed information about the accuracy of the
underlying dipole calculation.

Within the same dipole formalism one can calculate charm production in diffractive 
DIS, $l+p\to l'+\bar cc+p$. This cross section turns out to be a higher twist, $\sim 
1/m_c^4$. However, semi-inclusive diffractive production, $l+p\to l'+\bar ccX+p$, is a 
leading twist, $\sim 1/m_c^2$ due to additional gluon radiation.
It was demonstrated recently \cite{kst-hf} that diffractive hadro-production of heavy 
flavors is also a leading twist, and dipole model calculations well explain available 
data.

\section*{Acknowledgment} This work was partially supported by BMBF (Germany) grant
06HD158; by DFG (Germany)  grant PI182/3-1; and by Fondecyt (Chile) grant 1050519.

\appendix  
\section{Experimental cuts}  
The experimental cuts are given with respect to the electron beam  
line. We relate the transverse momenta relative to the electron beam  
line and the virtual photon.  
\begin{figure}[htbp]  
\begin{center}  
\resizebox{6cm}{!}{\includegraphics{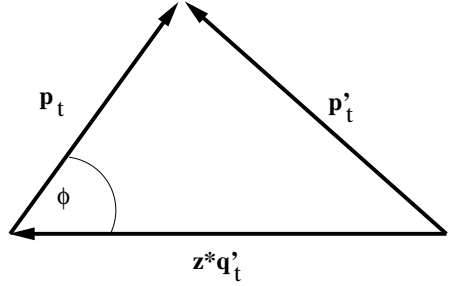}}  
\caption{Relating transverse momenta relative to the electron beam  
  line, $p_{D}^{\bot'}$, and to the virtual photon direction  
  $p_{D}^{\bot}$ for a small rotation of the z axis.}  
\label{fig:ptslash}  
\end{center}  
\end{figure}  
As shown in Fig.~\ref{fig:ptslash} we can write  
\begin{eqnarray}  
\vec p_{D}^{\bot'}&=&\vec p_{D}^{\bot}+\frac{z}{\alpha}\vec q_{t}'\nonumber\\  
&=&\left(\begin{array}{c} p^{x}_{D}\\p_{D}^{y} \end{array}\right)  
+\left(\begin{array}{c} -\frac{z}{\alpha}q_{t}'\\0 \end{array}\right)  
\label{eqn:vecp}  
\end{eqnarray}  
and therefore  
\begin{equation}  
p_{D}^{\bot'}=\sqrt{\left(\frac{z}{\alpha}\,q_{t}'\right)^{2}+  
(p^{\bot}_{D})^{2}-2\,\frac{z}{\alpha}\,q_{t}'\,p_{D}^{\bot}\,\cos{\phi}}\ ,  
\end{equation}  
where the transverse virtual photon momentum $q_{t}'$ is given by kinematics  
\begin{eqnarray}  
q_{t}'&=&\frac{Q^{2}M}{s}\sqrt{\frac{s^{2}}{Q^{2}\,M^{2}}(1-y)-1}\\  
&\simeq&Q\sqrt{1-y}.  
\label{eqn:qt}  
\end{eqnarray}  
The momenta are shown in Fig.~\ref{fig:ptslash}.   
The quantities relative to the electron direction are indicated with a prime.   
The transverse photon momentum is accompanied by a factor $z/\alpha=z^{*}$.   
This factor evolves because the virtual photon splits into a heavy quark pair and   
not directly into $D$ mesons.\\


\begin{thebibliography}{99}  

\bibitem{zkl} B.Z.~Kopeliovich, L.I.~Lapidus and A.B.~Zamolodchikov, Sov. Phys.
JETP Lett. {\bf 33}, 595 (1981); Pisma v Zh. Exper. Teor. Fiz. {\bf 33}, 612
(1981).

\bibitem{Levelt:1993ac}  
  J.~Levelt and P.J.~Mulders,  
  Phys.\ Rev.\ D {\bf 49}, 96 (1994)  
  [arXiv:hep-ph/9304232].  
  

\bibitem{le}  
J. Levelt, Deep inelastic semi-inclusive processes,  PhD thesis, 1993.  
 

\bibitem{be} E. Berger, Proc. Workshop on electronuclear physics with  
internal targets, SLAC, eds. R.G. Arnold and R. Minehart, (1987).  

\bibitem{he}  
S.J. Brodsky, A. Hebecker and E. Quack, Phys. Rev. D {\bf 55}, 2584 (1997).  

\bibitem{bj}  
J.D. Bjorken, J.B. Kogut and D.E. Soper, Phys. Rev. D {\bf 3}, 1382 (1971).  

\bibitem{kst1}
B.Z. Kopeliovich, A. Schaefer and A. V. Tarasov, Phys. Rev. C {\bf 59}, 1609 (1999).

\bibitem{hir}
B.Z. Kopeliovich, proc.\ of the workshop Hirschegg '95:
Dynamical Properties of Hadrons in Nuclear Matter, Hirschegg January
16-21, 1995, ed. by H. Feldmeyer and W. N\"orenberg, Darmstadt, 1995,
p. 102 [arXiv:hep-ph/9609385].

\bibitem{Shoshi:2002in}  
  A.I.~Shoshi, F.D.~Steffen and H.J.~Pirner,  
  Nucl.\ Phys.\ A {\bf 709}, 131 (2002)  
  [arXiv:hep-ph/0202012].  

\bibitem{gbw}  
K. Golec-Biernat, M. W\"usthoff, Phys. Rev. D {\bf 59}, 014017 (1999) [arXiv:hep-ph/9807513].  

\bibitem{dis}  
R. Devenish, A. Cooper-Sarkar, "Deep Inelastic Scattering", Oxford University Press 2004.  

\bibitem{pe}  
C. Peterson, D. Schlatter, I. Schmitt and P.M. Zerwas, Phys. Rev. D {\bf 27}, 105 (1983).  

\bibitem{h1}  
H1 Collab., A. Aktas et al., Eur. Phys. J. C {\bf 38}, 447-459, 08/04 (2005).  


\bibitem{Yao:2006px}  
  W.M.~Yao {\it et al.}  [Particle Data Group],  
  J.\ Phys.\ G {\bf 33}, 1 (2006).  
  

\bibitem{pdg}  
S. Eidelmann et al., Phys. Lett. B {\bf 592}, 1 (2004).  

\bibitem{h1alt}  
H1 Collaboration, C. Adloff et. al., Nucl. Phys. B {\bf 545}, 21-44 (1999) [arXiv:hep-ex/9812023].  
 

\bibitem{Kruger} M. Krueger, private communication.

\bibitem{Bartels:2002cj}  
  J.~Bartels, K.~Golec-Biernat and H.~Kowalski,  
  Phys. Rev. D {\bf 66}, 010001 (2002)  
  [arXiv:hep-ph/0203258].  

\bibitem{Kopeliovich:2007yv}
  B.Z.~Kopeliovich, A.H.~Rezaeian, H.J.~Pirner and I.~Schmidt,
  arXiv:0704.0642 [hep-ph].
  


\bibitem{Shoshi:2002fq}
  A.I.~Shoshi, F.D.~Steffen, H.G.~Dosch and H.J.~Pirner,
  Phys.\ Rev.\  D {\bf 66}, 094019 (2002)
  [arXiv:hep-ph/0207287].
 

\bibitem{kt} B.Z. Kopeliovich and A.V.~Tarasov, Nucl. Phys. A {\bf 710}, 180 (2002). 

\bibitem{Kopeliovich:2003cn}  
  B.Z.~Kopeliovich and J.~Raufeisen,  
  Lectures given at International School on Heavy Quark Physics, 
  Dubna, Russia, 27 May - 5 Jun 2002 [arXiv:hep-ph/0305094].  

\bibitem{kst-hf} B.Z. Kopeliovich, I.~Schmidt and A.V.~Tarasov
[arXiv: hep-ph/0702106].


\end{thebibliography}
\end{document}